\documentclass[useAMS,usenatbib]{mn2e}
\usepackage{psfig}
\usepackage{epsf}
\newcommand{\ltaraw}{$\; \buildrel < \over \sim \;$}
\newcommand{\lta}{\lower.5ex\hbox{\ltaraw}}
\newcommand{\gtaraw}{$\; \buildrel > \over \sim \;$}
\newcommand{\gta}{\lower.5ex\hbox{\gtaraw}}

\newcommand{\kms}{{\rm\,km\,s^{-1}}}

\newcommand{\msun}{{\rm\,M_\odot}}

\newcommand\ion[2]{#1$\;${\small{#2}}\relax}%

\loadboldmathitalic 
\title [Microlensing of Fractal BLRs]
{Gravitational microlensing of quasar Broad Line Regions:  
the influence of fractal structures} 
\author[Geraint F. Lewis \& Rodrigo A. Ibata]
{Geraint F. Lewis$^1$ \& Rodrigo A. Ibata$^2$\\
$^1$Institute of Astronomy, School of Physics,
University of Sydney, NSW 2006, Australia: {\tt gfl@physics.usyd.edu.au} \\
$^2$Observatoire de Strasbourg, 11 Rue de l'Universite,F-6700 Strasbourg, France: 
{\tt ibata@astro.u-strasbg.fr}\\
}
\date{\today}
\begin{document} 
\maketitle 
\begin{abstract}
Recent models for the emission  clouds within the Broad Line Region of
quasars suggest that they are due to transient overdensities within an
overall  turbulent medium.   If this  were  the case,  the broad  line
emission  would spatially  appear fractal,  possessing structure  on a
range of  scales. This  paper examines the  influence of  such fractal
structure when a quasar is  microlensed by a population of intervening
masses. It is found that  while the highest fractal levels can undergo
significant  microlensing magnification, when  these light  curves are
superimposed  to  create  an  emission  line  profile,  the  resultant
emission line profile remains  relatively constant for physical models
of the  Broad Line Region. It  is concluded that the  detection of the
possible  fractal structure  of Broad  Line Regions  via gravitational
microlensing is not practical.
\end{abstract}

\section{Introduction}\label{introduction}
While quasars are  amongst the most luminous objects  in the Universe,
the  majority of  their emission  is  generated within  a region  only
parsecs in extent. At cosmological scales, such regions are well below
the resolving power  of even the most powerful  telescopes, and so the
spatial  structure  within  these  inner regions  cannot  be  directly
observed and must be inferred by other, more indirect means.

In  recent years,  a general  picture for  quasar central  regions has
emerged, with  a hot accretion  disk surrounding a  supermassive black
hole   being  responsible   for  the   extensive   continuum  emission
characteristic  of quasars. Beyond  this, on  the scale  of $\sim1pc$,
this  accretion disk  is surrounded  by a  large population  of clouds
which  are illuminated  by the  central engine  and produce  the broad
emission   lines,   with  widths   of   thousands   of  $\kms$,   also
characteristic  of  quasar spectra\footnote{It  should  be noted  that
alternative models  for the source  of the broad line  emission exist,
including the accretion disk wind model of \citet{1995ApJ...451..498M}
which occurs on a considerably smaller scale.}.

Initial studies  of the Broad Line  Region (BLR) were  based on simple
ionisation models, with predicted BLR sizes of 0.1 to a few parsecs in
extent \citep[e.g.][]{1979RvMP...51..715D}.  Recent studies, utilising
reverberation mapping to  measure the scale of the  BLR, however, have
found  evidence for a  smaller BLR,  more than  an order  of magnitude
smaller       than      suggested      by       ionisation      models
\citep{1985ApJ...298..283P}.  Furthermore, \citet{1999ApJ...526..579W}
and \citet{2000ApJ...533..631K} have demonstrated that the size of the
BLR scales with quasar  luminosity, with $R_{BLR}\propto L^{0.7}$, and
BLRs were also found to possess significant ionisation stratification,
with high-ionisation lines  arising in a region an  order of magnitude
smaller than low-ionisation lines.

The  origin of  the BLR  has  proved problematic  to understand,  with
suggestions  that  the emitting  material  could  be  the debris  from
star-disk    collisions     at    the    heart     of    the    quasar
\citep{1994ApJ...434...46Z,1996ApJ...470..237A},  supernova explosions
within                         quasar                         outflows
\citep{2001A&A...375..827P,2003A&A...408...79P},  or even  the bloated
atmospheres of irradiated stars \citep{1997MNRAS.284..967A}.  However,
the  BLR must  possess  significant substructure,  with line  emission
arising  from  dense  clouds   embedded  in  a  lower  density  medium
\citep{1981ApJ...245..396C},  with the  smoothness  of emission  lines
suggesting  that  there  must  be  of  order  $10^8$  emitting  clouds
\citep{1998MNRAS.297..990A}, effectively ruling out the picture of the
BLR being  composed of  a relatively small  number of large  clouds or
bloated   stars.   Nevertheless,  such   a  situation   is  physically
problematic, as  these clouds  should rapidly dissipate,  with various
mechanisms  suggested  for their  confinement  [e.g.  magnetic  fields
\citet{1987MNRAS.228P..47R}].

Recently,  \citet{2001ApJ...549..118B}  suggested  that  the  `clouds'
within the BLR are not isolated, individual entities embedded within a
confining  medium, but  rather represented  transient knots  of higher
density within an overall turbulent  BLR. One conclusion of this study
was  that the  BLR would  not  appear as  a smoothly-varying  emitting
region. Instead, the  turbulent nature of the BLR  would result in the
spatial  emission  from  the  region  possessing  an  overall  fractal
structure.

Gravitational  microlensing  provides  an  opportunity  to  `see'  the
structure in  the central  regions of quasars,  utilising differential
magnification    effects    \citep{1991AJ....102..864W}.     Recently,
\citet{fractal}  demonstrated   that  small-scale  fractal  structures
within  quasars, arising  in an  x-ray emitting  hot corona  above the
accretion disk, imprint themselves on the light curve of a microlensed
quasar.  Hence, this paper  considers the influence of microlensing on
an extended, fractal BLR.  However, as this is more extensive than the
X-ray emitting  region, it is important  to compare it  to the natural
scale length for gravitational  microlensing; the Einstein radius.  In
the source plane, this is given by
\begin{equation}
\eta = \sqrt{ 4 \frac{ G M }{ c^2 } \frac{D_{ls} D_{os}}{D_{ol}}}
\label{eqn1}
\end{equation}
where  $M$ is  the mass  of the  microlensing body,  and  $D_{ij}$ are
angular diameter distances between  the observer $(o)$, lens $(l)$ and
source  $(s)$. For typical  cosmological lensing  configurations, with
microlensing  stars with  mass $M\lta  1\msun$, this  scale  length is
$\eta\lta 0.1{\rm pc}$, and  is often substantially smaller. The X-ray
emitting  region considered  in \citet{fractal}  is small  compared to
this  scale  and   hence  significant  microlensing  magnification  is
expected.  As  discussed in \citet{2004MNRAS.348...24L},  however, the
extensive nature of  the BLR ensures it must  lie across a substantial
portion of  the complex caustic network  that is seen  at high optical
depths, and the effects of  microlensing cannot be approximated by the
influence  of individual  caustic structures.   The structure  of this
paper  is  as follows;  Section~\ref{method}  describes the  numerical
approach adopted in  this study, while Section~\ref{results} discusses
the  resultant   microlensing  light  curves   and  their  statistical
properties.   The   conclusions  to   this  study  are   presented  in
Section~\ref{conclusions}.

\begin{figure}
\centerline{ \psfig{figure=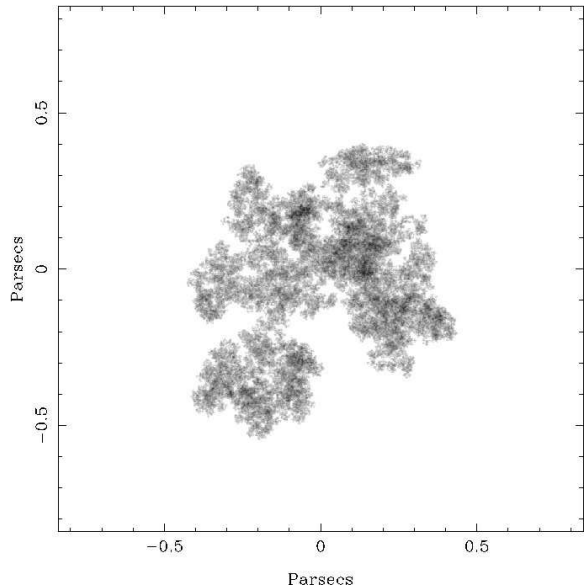,angle=270,width=3.0in}}
\caption[]{A   realisation  of   a   fractal  BLR   as  described   in
\citet{2001ApJ...549..118B}. Clearly clumps and subclumps can be seen,
but the smallest structures  in this picture ($\sim10^{-6}$pc) are too
small to be seen by eye.
\label{fig1}}
\end{figure}

\section{Method}\label{method}

\subsection{Cloud Distributions}\label{clouds}
The argument that BLRs possess  fractal structure was expounded by the
turbulence  model  of  \citet{2001ApJ...549..118B},  and  it  is  this
procedure for  defining a fractal cloud distribution  that is utilised
in this paper. The fractal structure occupies a region with an overall
scale  size of $R_{max}$.  Within this  radius there  are a  series of
hierarchies whose radius is given by
\begin{equation}
R(h) = R_{max}\ L^{-h}
\label{hier}
\end{equation}
where $h = 0,  1 \cdots H$ is the level of  the hierarchy, $H$ denotes
the maximum fractal hierarchy in  the structure and $L$ is a geometric
factor.  Each hierarchy possesses $N$ substructures (the multiplicity)
and the overall fractal dimension is defined to be
\begin{equation}
D = \frac{ \log{ N } }{ \log{ L } }\ .
\end{equation}
To  generate a fractal  distribution, a  region $R_{max}$  is defined;
this is  the zeroth  level of the  hierarchy with $h=0$.   Within this
region, $N$ locations  are chosen at random and  represent the centres
of the first level of the hierarchy (with $h=1$) and assigned a radius
given  by  Equation~\ref{hier}. Within  each  individual subregion,  a
further $N$ centres are scattered  at random and are assigned a radius
corresponding to  the second level  of the hierarchy, with  $h=2$. The
process is continued  until the highest level of  the hierarchy, $h=H$
is reached; each point in this upper-most hierarchy corresponds to the
location of a cloud with  a radius given by Equation~\ref{hier}.  This
procedure  results  in  a  total  of  $n_{tot}  =  N^H$  clouds  being
distributed over  a region.  For the  purposes of this  study, each of
these final clouds was assumed to have the same luminoisty, although a
more  realistic  model  would   probably  require  a  distribution  of
luminosities throughout the region.

In     considering     a    realistic     model     for    the     BLR
\citet{2001ApJ...549..118B} determined the fractal parameters required
to  reproduce known  physical properties  of the  region, such  as its
column density and covering factor.  These values are employed in this
current study, with  $R_{max} = 5.12 R_{BLR}$, where  $R_{BLR}$ is the
nominal  radius   to  the  BLR,  $L=3.2$  and   $N=14$  (see  footnote
2)\footnotetext{In    the   study    of   \citet{2001ApJ...549..118B},
$N=14.62$, and  fractal structures could  be generated by  drawing the
number  in a  particular hierarchy  with  this being  the mean  value.
Fixing $N=14$  does not significantly influence  the resulting fractal
structure.}. The  maximum hierarchy which will be  employed is related
to  the resolution  of the  microlensing magnification  map,  which is
discussed in Section~\ref{lensing}.

In choosing $R_{BLR}$,  the scale relation between the  BLR radius and
the luminosity of the quasar, as derived from reverberation mapping is
employed      \citep{1999ApJ...526..579W,2000ApJ...533..631K}.      As
discussed  in  the  next  section,   this  study  will  focus  on  the
microlensed images of the quasar Q2237+0305, which possess an absolute
magnitude of  $M\sim-26$ [accounting  for a magnification  of $\sim16$
\citep{1998MNRAS.295..488S}]; as  shown in \citep{2004MNRAS.348...24L}
this  corresponds to a  $R_{BLR}\sim0.05{\rm pc}$  for high-ionization
emission  (e.g.   \ion{C}{IV}   )  and  $R_{BLR}\sim0.4{\rm  pc}$  for
low-ionisation emission  (e.g.  \ion{Mg}{II}  ).  For the  purposes of
this study,  $R_{BLR}=0.1{\rm pc}$, a  value between the high  and low
emission scale  sizes.  Using these parameters, a  typical fractal BLR
can be generated, with an example presented in Figure~\ref{fig1}.

\subsection{Gravitational Microlensing}\label{lensing}
As noted  in Section~\ref{introduction},  the substantial size  of the
BLR implies that the influence of gravitational microlensing cannot be
approximated  as   a  simple  point-mass  lens   or  isolated  caustic
structure.   To  this  end,   the  backward  ray-tracing  approach  of
\citet{1986A&A...166...36K}  and \citet{1990ApJ...352..407W}  was used
to generate the magnification map employed in this study.

\begin{figure}
\centerline{ \psfig{figure=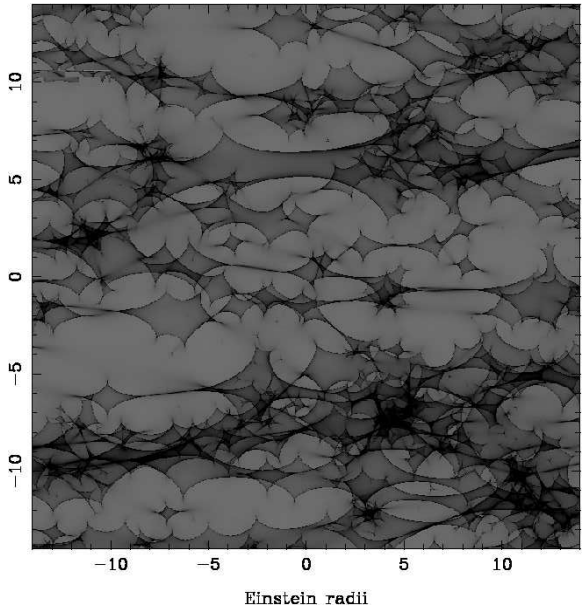,angle=270,width=3.0in}}
\caption[]{The magnification map  employed in this study. Representing
image A of the microlensed quasar, Q2237+0305, a dimensionless surface
density  of $\kappa=0.36$  and  external shear  of $\gamma=0.41$  were
employed. The  units are in Einstein  radii for a solar  mass star and
the scale of the image matches that presented in Figure~\ref{fig1}.
\label{fig2}}
\end{figure}

The  macrolensing  parameters were  chosen  to  represent  Image A  of
Q2237+0305,  the most intensively-studied  microlensed quasar,  with a
dimensionless surface  mass density  of $\kappa=0.36$ and  an external
shear  of  $\gamma=0.41$;  this  surface mass  density  describes  the
focusing due to matter within the beam of light traversing the galaxy,
whereas the  shear is  the distortion  of the beam  due to  the larger
scale distribution of  matter \citep{1986A&A...166...36K}.  All of the
mass is distributed in microlenses with masses of $1\msun$. Given that
the   lensing  galaxy  in   this  system   possesses  a   redshift  of
$z_l=0.0395$, while the source  quasar is at $z_s=1.695$, the Einstein
Radius  (Equation~\ref{eqn1}) for a  solar mass  star is  $\eta_o \sim
0.06\rm{pc}$,   adopting  the   current  concordance   cosmology.   To
accommodate  the  model  BLR  described in  Section~\ref{clouds},  the
entire  ray  tracing   region  is  taken  to  be   28  Einstein  radii
$(\sim1.68{\rm pc})$ on a  side.  The resulting magnification map used
in this study is presented in Figure~\ref{fig2}.

\begin{figure*}
\centerline{ \psfig{figure=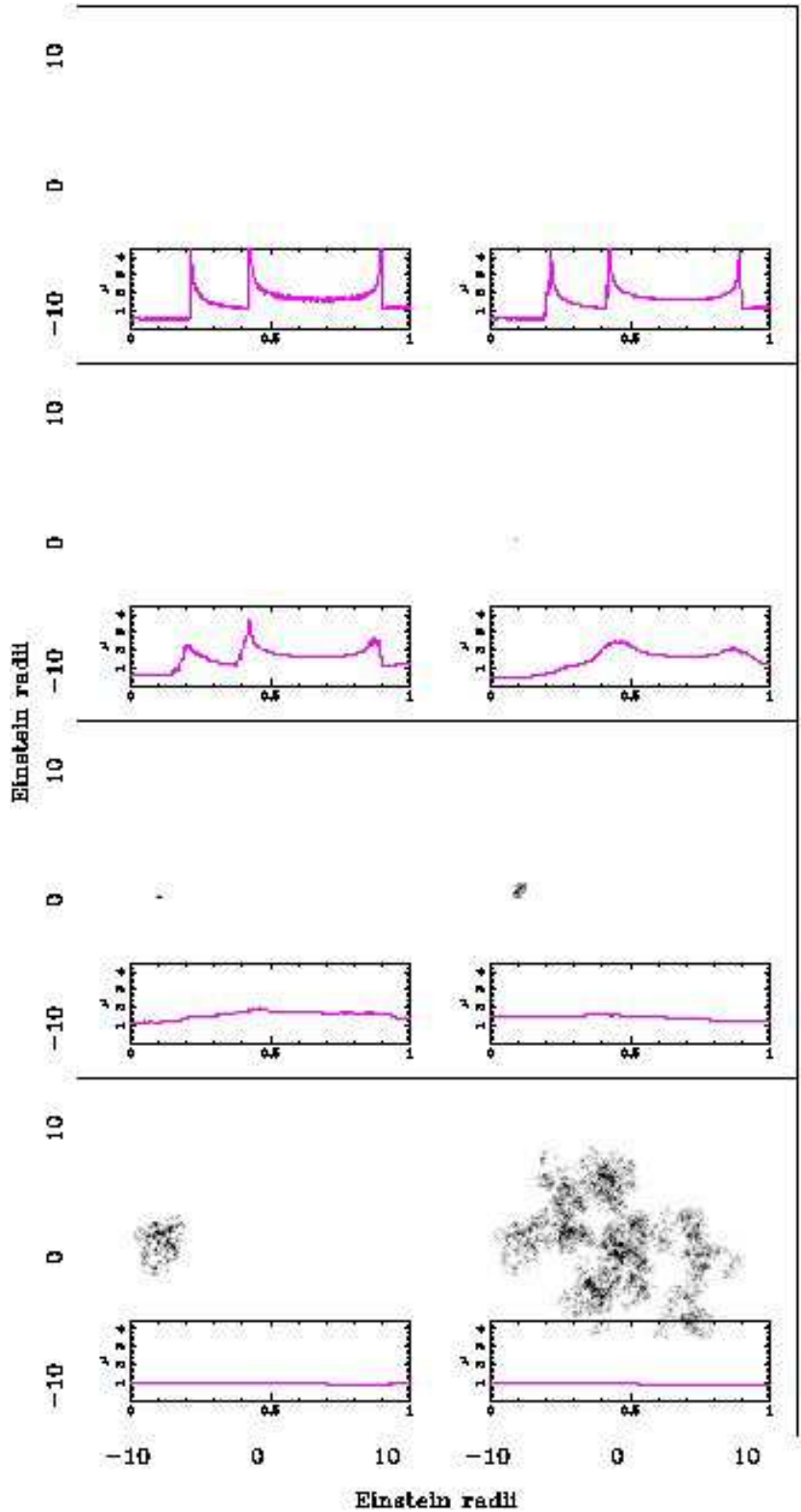,angle=270,width=4.3in}}
\caption[]{Example light curves  for the fractal hierarchies presented
in  this paper,  from the  smallest  ($h=7$ top-left)  to the  largest
($h=0$ bottom-right). Each panel presents the cloud distribution (note
that the  smallest structures  are not visible  on the scale  of these
figures),   with  the   resulting   light  curve   arising  from   the
magnification map presented in Figure~\ref{fig2}.  Each light curve is
1 Einstein  radii in extent,  with the location  of the clouds  in the
figure representing  the centre of the  position at the  centre of the
light curve.
\label{fig3}}
\end{figure*}

\subsection{Image pixelation}\label{hierarchy}
Clearly, the  fractal clouds outlined  in Section~\ref{clouds} possess
structure  on a  range of  scales and  this must  be reflected  in the
magnification map. Hence the  magnification map utilised in this study
possesses a pixel  scale of 1000 pixels per  Einstein Radius, with one
pixel  physically corresponding to  $6\times10^{-5}{\rm pc}$,  and the
total map  presented in Figure~\ref{fig2}  is 28000 pixels on  a side.
If a pixel  in this map represents a single cloud  in the fractal BLR,
then the maximum hierarchy  in the fractal distribution corresponds to
$H\sim7$   and  the   total   number   of  clouds   in   the  BLR   is
$N_{tot}\sim1.5\times10^{9}$.      Note    that    the     model    of
\cite{2001ApJ...549..118B} considered  a maximum fractal  hierarchy of
$H=11$, with a  total of $N_{Tot}\sim4\times10^{14}$ individual clouds;
with this, the scale  of each cloud would be $\sim1.4\times10^{-6}{\rm
pc}$ and  the magnification  map presented in  Figure~\ref{fig2} would
have to be $1.2\times10^6$ pixels  along each side, a very challenging
prospect. As  discussed later, the  lower resolution map  adopted here
does not significantly influence the conclusions of this paper.

\subsection{Generating light curves}\label{lightcurves}
In typical gravitational microlensing problems, magnification maps are
convolved with  the surface brightness profile of  a preferred source,
providing  a map  from which  the microlensing  light curve  from that
source can  be drawn. Such a convolution,  utilising Fourier transform
techniques, was  found to be computationally too  expensive, given the
number of pixels used in the magnification map. Hence, in this present
study,  light curves  were  simply generated  by  extracting a  single
column of  pixels along a magnification  map at the location  of a BLR
cloud.   Summing each  of these  columns  for a  collection of  clouds
together  then gives  the overall  light curve  of the  BLR  region of
interest. With this approach,  each individual BLR cloud (the smallest
structure within the  entire BLR) has a size of a  single pixel in the
map, corresponding to $6\times10^{-5}{\rm pc}$.

\section{Results}\label{results}

Figure~\ref{fig3} presents a series  of light curves derived from this
analysis. Each  panel presents  subsections of the  fractal hierarchy,
from $h=7$ in  the top-left, consisting of a total  of 14 sources), to
$h=0$ in the bottom-right, with a total of $1.5\times10^9$ clouds. The
background  of each  panel presents  a  greyscale map  of the  fractal
structure (note in $h=7$ and $h=6$ the structure is too small to see).
Each subpanel  presents the microlensing  light curve for  the fractal
hierarchy;  the x-axis  is in  units  of Einstein  radii, whereas  the
y-axis corresponds  to the magnification of the  region.  The expected
crossing  time  of an  Einstein  radius  in  this system  (assuming  a
transverse  velocity of  the  lens of  $\sim600$km/s)  is $\sim8$  yrs
\citep[e.g.][]{1996MNRAS.283..225L}.    The   degree  of   variability
changes  quite strongly  with fractal  hierarchy $h=7$  displaying the
strong variability, with caustic crossings quite similar to point-like
sources\footnote{The slight fuzz visible in  the light curve is due to
low-level  numerical noise  in the  magnification maps.   Its presence
does not influence the results  of this study.}.  Clearly these source
sizes at  fractal hierarchies of $h>7$ are  substantially smaller than
the  fractal  structure  seen   in  Figure~\ref{fig2}  and  hence  the
resulting  light curves  for individual  sources at  this  level would
appear qualitatively  the same  as the $h=7$  case, although  the peak
magnifications would  be larger.  Considering lower values  of $h$, it
can be  seen that  the light curve  fluctuations are smoothed  out, as
expected  as  the  source size  increases~\citep{1991AJ....102..864W},
while at  the base fractal  level ($h=0$), which considers  the entire
cloud population, the light curve is quite flat.

\begin{figure}
\centerline{ \psfig{figure=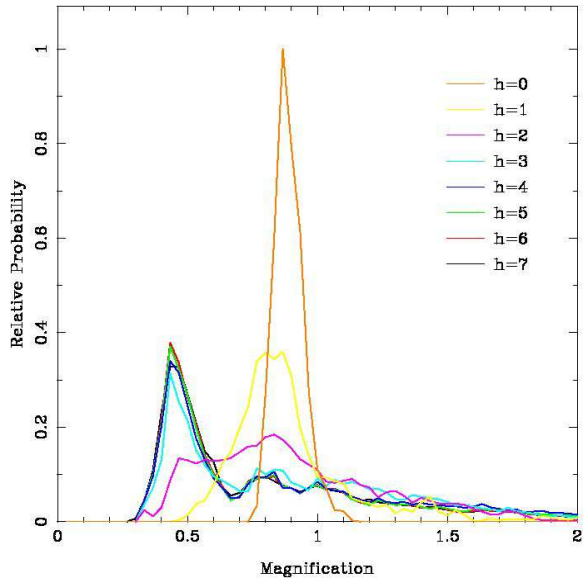,angle=270,width=3.0in}}
\caption[]{The magnification distributions for 250 realisations of
fractal BLRs combined with the magnification map displayed in
Figure~\ref{fig2}.
\label{fig4}}
\end{figure}

To examine this  further, 250 light curve samples  were generated with
differing  realisations of  the fractal  hierarchy.  Figure~\ref{fig4}
presents the  magnification probability distributions for  each of the
fractal hierarchies in the entire sample. These confirm the properties
of the light curve seen in Figure~\ref{fig3}, with the highest fractal
hierarchy displaying  magnifications of $10-20\%$.   This distribution
is  quite different  to the  lower fractal  hierarchies  which display
typical  light  curves  for  small  sources.  Here  the  magnification
probability distribution is dominated by  a peak at a magnification of
0.5,  corresponding to  extensive  periods of  demagnification in  the
light curve,  with a high magnification tail  representing the periods
of caustic crossing~\citep{1992ApJ...386...19W,1995MNRAS.276..103L}.

A simple interpretation of this  result, therefore, is that if BLRs do
indeed  possess the fractal  structure outlined,  then, on  the whole,
they   should  not   be  significantly   magnified   by  gravitational
microlensing. While this is true, what it actually implies is that the
total  flux in  the  emission line  will  not vary  as  the quasar  is
microlensed,  but  it must  be  remembered  that  as well  as  spatial
structure, BLRs also possess  kinematic structure which is responsible
for  the  width   of  the  emission  line  [see   earlier  studies  by
\citet{1988ApJ...335..593N} and \citet{1990A&A...237...42S}].

\begin{figure}
\centerline{ \psfig{figure=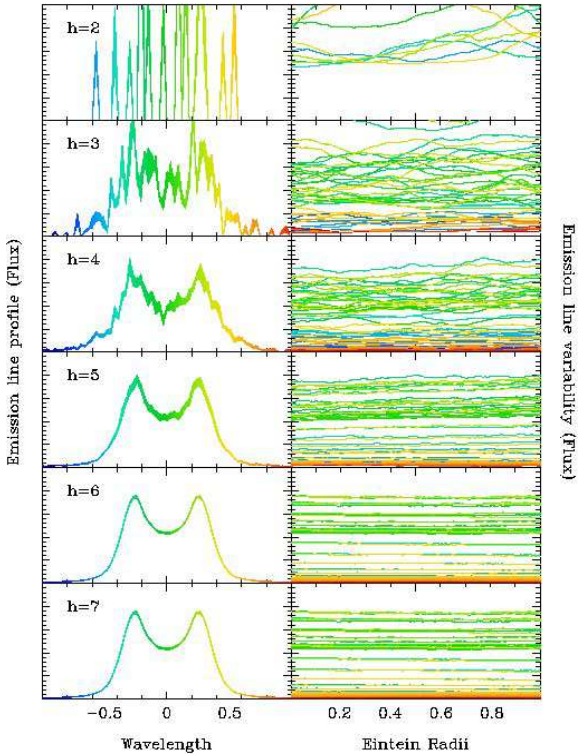,angle=270,width=3.0in}}
\caption[]{The  emission line  profile  for the  toy kinematic  models
presented in  this paper.  The  left-hand panel presents  the emission
line profile for the appropriate bulk motion of the fractal structure.
The right-hand panel presents the  light curve for various portions of
the  lines, colour-coded  to match  the wavelengths  presented  in the
left-hand panels.  In this figure,  the total fractal hierarchy is the
same for all  frames and the $h$ values denote  the hierarchy at which
the bulk velocity has been assigned; i.e. in the top panel with $h=2$,
each $h=2$ has been assigned a  velocity drawn from the model and this
value has been assigned to each higher fractal level within that $h=2$
structure.  Hence  the $h=7$  possesses more kinematic  diversity that
$h=2$, resulting  in the overall smoothness of  the resulting emission
line.
\label{fig5}}
\end{figure}

As  noted  in \citet{2001ApJ...549..118B},  however,  the fractal  BLR
model they  present is  purely geometrical and  does not  consider the
kinematic aspects  of the BLR. For  the purposes of this  paper, a toy
model for  the kinematic  properties is considered,  with clouds  of a
particular  fractal  hierarchy  assigned   a  velocity  drawn  from  a
Keplerian distribution  with random projections (i.e. all  clouds at a
particular fractal hierarchy and above are assigned the same projected
velocity). This  velocity structured BLR was then  microlensed and the
light curves summed into kinematic bins.  The results of this exercise
are presented  in Figure~\ref{fig5}; the left-hand  panels present the
emission line  profile, comprised  of 1000 velocity  elements, whereas
the right-hand panel  presents the light curves for  sampled values of
the velocity  elements over  one Einstein radius.   The spread  in the
emission  line profile  represents the  superposition of  all emission
lines along  the one  Einstein radius.  The  pairs of  panels arranged
vertically  represent the  the various  velocity models;  i.e.  in the
$h=7$  model, each  of  the  $14$ clouds  within  the highest  fractal
hierarchy are assigned the  same Keplerian velocity, for $h=6$, $14^2$
within the  second highest are  assigned the same  Keplerian velocity,
etc.  Clearly,  the $h=7$ emission  line profile displays  very little
variability.  Even at this high resolution, each velocity element must
be the sum of of the light curves of many small sources, averaging out
to  give an  overall flat  light curve.   In moving  to  lower fractal
hierarchies,  the  emission  line   profiles  display  more  and  more
variability.  However,  it is  clear that bulk  motions of  the lowest
fractal  hierarchies result  in  very ``choppy''  structure, with  the
$h=2$  emission line  breaking  down  into a  series  of spikes.   The
smoothness of observed emission lines suggest that such a situation is
unphysical \citep[e.g.][]{1998MNRAS.297..990A} and hence the kinematic
properties must be  defined by the higher fractal  hierarchies. It can
be concluded, therefore, that  the fractal structure will be virtually
impossible  to  detect  by  studying  the emission  line  profiles  of
microlensed quasars.

\section{Conclusions}\label{conclusions}
The nature of the broad  line emission regions of quasars has remained
a  subject of discussion  for a  number of  years, with  recent models
suggesting that the emitting clouds are temporary density enhancements
in an overall  turbulent medium.  Such a scenario  predicts that these
density enhancements should have a fractal distribution of structures.
This   paper   has  investigated   the   influence  of   gravitational
microlensing  on  this   extensive  fractal  structure,  finding  that
different   levels  of   the  fractal   structure   undergo  differing
magnifications,  with   the  smaller  sections   of  the  substructure
suffering stronger  magnification. It  was found, however,  that while
individual clouds  in the  fractal hierarchy displayed  quite dramatic
variability due to microlensing,  the combination of light curves from
the entire population of clouds  within the BLR resulted in an overall
constant light curve.

While the  overall magnification  of the BLR  appears to  be constant,
more significant magnification of substructures could be apparent when
examining the  form of the  emission line profile,  i.e. substructures
within  the  BLR  could  possess  coherent velocities  and  hence  may
contribute  to a particular  velocity within  the emission  line. This
paper  considered  a  simple  Keplerian  velocity  structure  for  the
emission line  clouds, assigning a particular velocity  to all fractal
hierarchies below  a particular level.  The result  of this procedure,
however, revealed that coherent velocity structure cannot apply to the
lowest levels of the fractal  hierarchy as the resultant emission line
profile is  clearly too  structured, although dramatic  variability is
seen throughout the line. Assigning coherent velocity structure to the
higher  fractal hierarchy  does smooth  out the  form of  the emission
line, but it also smooths out the light curves for each velocity bin.
From this it is possible  to conclude that while individual structures
within  a BLR  are being  microlensed and  are  undergoing significant
magnifications, the  extensive nature of  the BLR and  the requirement
that the resulting emission  line appears relatively smooth means that
microlensing is  unlikely to reveal the putative  fractal structure of
quasar BLRs.


\newcommand{\aap}{A\&A} 
\newcommand{\apj}{ApJ} 
\newcommand{\apjl}{ApJ}
\newcommand{\aj}{AJ}                         
\newcommand{\mnras}{MNRAS}
\newcommand{\apss}{Ap\&SS} 
\newcommand{\nat}{Nature}


\begin{thebibliography}{DUM}
\bibitem[\protect\citeauthoryear{Abajas et al.}{2002}]{2002ApJ...576..640A} 
Abajas C., Mediavilla E., Mu{\~ n}oz J.~A., Popovi{\' c} L.~{\v C}., Oscoz 
A., 2002, ApJ, 576, 640 

\bibitem[\protect\citeauthoryear{Alexander \& 
Netzer}{1997}]{1997MNRAS.284..967A} Alexander T., Netzer H., 1997, MNRAS, 
284, 967 

\bibitem[\protect\citeauthoryear{Arav et al.}{1998}]{1998MNRAS.297..990A} 
Arav N., Barlow T.~A., Laor A., Sargent W.~L.~W., Blandford R.~D., 1998, 
MNRAS, 297, 990

\bibitem[\protect\citeauthoryear{Armitage, Zurek, \& 
Davies}{1996}]{1996ApJ...470..237A} Armitage P.~J., Zurek W.~H., Davies 
M.~B., 1996, ApJ, 470, 237 

\bibitem[\protect\citeauthoryear{Bottorff \& Ferland}{2001}]{2001ApJ...549..118B} 
Bottorff M., Ferland G., 
2001, \apj,  549, 118

\bibitem[\protect\citeauthoryear{Capriotti, Foltz, \& 
Byard}{1981}]{1981ApJ...245..396C} Capriotti E., Foltz C., Byard P., 1981, 
ApJ, 245, 396 

\bibitem[\protect\citeauthoryear{Davidson \& 
Netzer}{1979}]{1979RvMP...51..715D} Davidson K., Netzer H., 1979, RvMP, 51, 
715 

\bibitem[\protect\citeauthoryear{Kaspi et al.}{2000}]{2000ApJ...533..631K} 
Kaspi S., Smith P.~S., Netzer H., Maoz D., Jannuzi B.~T., Giveon U., 2000, 
ApJ, 533, 631 

\bibitem[\protect\citeauthoryear{Kayser et al.}{1986}]{1986A&A...166...36K} 
Kayser R., Refsdal S., Stabell R., 
1986, \aap,  166, 36

\bibitem[\protect\citeauthoryear{Lewis}{2004}]{fractal}
Lewis G.~F.,
2004, \mnras, 355, 106

\bibitem[\protect\citeauthoryear{Lewis \& 
Ibata}{2004}]{2004MNRAS.348...24L} Lewis G.~F., Ibata R.~A., 2004, MNRAS, 
348, 24 

\bibitem[\protect\citeauthoryear{Lewis \& 
Irwin}{1995}]{1995MNRAS.276..103L} Lewis G.~F., Irwin M.~J., 1995, MNRAS, 
276, 103 

\bibitem[\protect\citeauthoryear{Lewis \& 
Irwin}{1996}]{1996MNRAS.283..225L} Lewis G.~F., Irwin M.~J., 1996, MNRAS, 
283, 225 

\bibitem[\protect\citeauthoryear{Mediavilla et 
al.}{1998}]{1998ApJ...503L..27M} Mediavilla E., et al., 1998, ApJ, 503, L27 

\bibitem[\protect\citeauthoryear{Murray et al.}{1995}]{1995ApJ...451..498M} 
Murray N., Chiang J., Grossman S.~A., Voit G.~M., 1995, ApJ, 451, 498 
 
\bibitem[\protect\citeauthoryear{Nemiroff}{1988}]{1988ApJ...335..593N} 
Nemiroff R.~J., 
1988, \apj,  335, 593

\bibitem[\protect\citeauthoryear{Peitgen \& Saupe}{1988}]{wwe}
Peitgen, H.-O., Saupe, D., 1988, {\it The Science of Fractal Images}, 
Springer-Verlag (Berlin)

\bibitem[\protect\citeauthoryear{Peitgen \& Richter}{1986}]{awe}
Peitgen, H.-O., Richter, P.~H., 1986, {\it The Beauty of Fractals}, 
Springer-Verlag (Berlin)

\bibitem[\protect\citeauthoryear{Peterson, Crenshaw, \& 
Meyers}{1985}]{1985ApJ...298..283P} Peterson B.~M., Crenshaw D.~M., Meyers 
K.~A., 1985, ApJ, 298, 283 

\bibitem[\protect\citeauthoryear{Pittard et 
al.}{2003}]{2003A&A...408...79P} Pittard J.~M., Dyson J.~E., Falle 
S.~A.~E.~G., Hartquist T.~W., 2003, A\&A, 408, 79 

\bibitem[\protect\citeauthoryear{Pittard et 
al.}{2001}]{2001A&A...375..827P} Pittard J.~M., Dyson J.~E., Falle 
S.~A.~E.~G., Hartquist T.~W., 2001, A\&A, 375, 827 

\bibitem[\protect\citeauthoryear{Rees}{1987}]{1987MNRAS.228P..47R} Rees 
M.~J., 1987, MNRAS, 228, 47P 

\bibitem[\protect\citeauthoryear{Schmidt, Webster, \& 
Lewis}{1998}]{1998MNRAS.295..488S} Schmidt R., Webster R.~L., Lewis G.~F., 
1998, MNRAS, 295, 488 

\bibitem[\protect\citeauthoryear{Schneider \& Wambsganss}{1990}]{1990A&A...237...42S} 
Schneider P., Wambsganss J., 
1990, \aap,  237, 42

\bibitem[\protect\citeauthoryear{Wambsganss}{1992}]{1992ApJ...386...19W} 
Wambsganss J., 1992, ApJ, 386, 19 
 

\bibitem[\protect\citeauthoryear{Wambsganss \& 
Paczynski}{1991}]{1991AJ....102..864W} Wambsganss J., Paczynski B., 1991, 
AJ, 102, 864 

\bibitem[\protect\citeauthoryear{Wambsganss et al.}{1990}]{1990ApJ...352..407W} 
Wambsganss J., Paczynski B., Katz N., 
1990, \apj,  352, 407

\bibitem[\protect\citeauthoryear{Wandel, Peterson, \& 
Malkan}{1999}]{1999ApJ...526..579W} Wandel A., Peterson B.~M., Malkan 
M.~A., 1999, ApJ, 526, 579

\bibitem[\protect\citeauthoryear{Wayth, O'Dowd, \& Webster}{2005}]{Randall} 
Wayth, R.~B., O'Dowd, M., Webster, R.~L., 2005, \mnras, {\it Accepted, astro-ph/0502396}

\bibitem[\protect\citeauthoryear{Zurek, Siemiginowska, \& 
Colgate}{1994}]{1994ApJ...434...46Z} Zurek W.~H., Siemiginowska A., Colgate 
S.~A., 1994, ApJ, 434, 46 

\end{thebibliography}
\end{document}